Title: Novel magnetic-field-free switching behavior in vdW-magnet/oxide heterostructure

*Jihoon Keum, Kai-Xuan Zhang,*[$] Suik Cheon, Hyuncheol Kim, Jingyuan Cui, Giung Park, Yunyeong Chang, Miyoung Kim, Hyun-Woo Lee and Je-Geun Park**

Jihoon Keum, Dr. Kai-Xuan Zhang, Hyuncheol Kim, Jingyuan Cui, Giung Park, Prof. Je-Geun Park
Department of Physics and Astronomy, Seoul National University, Seoul 08826, South Korea
E-mail: kxzhang.research@gmail.com; jgpark10@snu.ac.kr

Jihoon Keum, Dr. Kai-Xuan Zhang, Hyuncheol Kim, Jingyuan Cui, Giung Park, Prof. Je-Geun Park
Center for Quantum Materials, Department of Physics and Astronomy, Seoul National University, Seoul 08826, South Korea

Dr. Kai-Xuan Zhang and Prof. Je-Geun Park
Institute of Applied Physics, Seoul National University, Seoul 08826, South Korea

Dr. Suik Cheon and Prof. Hyun-Woo Lee
Department of Physics, Pohang University of Science and Technology, Pohang 37637, South Korea

Yunyeong Chang and Prof. Miyoung Kim
Department of Materials Science & Engineering and Research Institute of Advanced Materials, Seoul National University, Seoul 08826, South Korea

Prof. Hyun-Woo Lee
Asia Pacific Center for Theoretical Physics, Pohang 37673, South Korea

[$]Present address: Department of Physics, Washington University in St. Louis, St. Louis, Missouri 63130, USA

Abstract: Magnetization switching by charge current without a magnetic field is essential for device applications and information technology. It generally requires a current-induced out-of-plane spin polarization beyond the capability of conventional ferromagnet/heavy-metal systems, where the current-induced spin polarization aligns in-plane orthogonal to the in-plane charge current and out-of-plane spin current. Here, we demonstrate a new approach for magnetic-field-free switching by fabricating a van-der-Waals magnet and oxide $Fe_3GeTe_2$/$SrTiO_3$ heterostructure. This new magnetic-field-free switching is possible because the current-driven accumulated spins at the Rashba interface precess around an emergent interface magnetism, eventually producing an ultimate out-of-plane spin polarization. This interpretation is further confirmed by the switching polarity change controlled by the in-plane initialization magnetic fields with clear hysteresis. We successfully combined van-der-Waals magnet and oxide for the first time, especially taking advantage of spin-orbit torque on the $SrTiO_3$ oxide. This allows us to establish a new way of magnetic field-free switching. Our work demonstrates an unusual perpendicular switching application of large spin Hall angle materials and precession of accumulated spins, and in doing so, opens up a new field and opportunities for van-der-Waals magnets and oxide spintronics.

**Main Text:** Van-der-Waals (vdW) magnets have come to offer a lot of promising aspects for both academic curiosity and potential applications [1-4]. The intrinsic two-dimensionality of the materials is a uniquely useful characteristic for any attempt to fabricate a working device on the nm scale or smaller. The other important factor to be considered is compatibility with materials already used in the industry. Therefore, it will be extremely important to be compatible with oxides, which have been used for decades in industry.

In-plane current-induced perpendicular magnetization switching under zero magnetic field is highly desired for next-generation Magnetic-Random-Access-Memory (MRAM) and in-memory computing. Over the past decades, various device designs have been proposed and tested to realize such magnetic-field-free switching [5-13]. Most of these studies have been conducted using MBE-grown ferromagnetic metals, but their high critical switching current remains an obstacle in realizing their practical application. The recent introduction of magnetic vdW materials to MRAM, especially for perpendicular magnetization switching with in-plane current injection, has gained increasing attention due to their two-dimensional (2D) magnetism [1,2,14-17] and energy-efficient switching [18-24]. Fe$_3$GeTe$_2$ (FGT) is a representative case of gigantic spin-orbit torque (SOT) in FGT-based heterostructure and FGT-alone systems [18-24]. However, both cases require a magnetic field to achieve deterministic switching, which is unfavorable in practical applications.

To realize magnetic-field-free switching, one needs an out-of-plane spin polarization [25], for example, produced by the current-driven SOT with certain symmetry constraints. In symmetry analysis, field-free switching requires broken symmetry of the mirror plane whose normal vector is perpendicular to both the out-of-plane and the charge current directions [26]. This condition allows for spin accumulation polarized in the perpendicular direction of a device and thus leads to deterministic perpendicular magnetization switching. Current-induced out-of-plane spin polarization using Weyl semimetals has been well studied in this scheme [6,25]. On the other hand, the out-of-plane spin polarization can be generated alternatively by a rare distinct mechanism, i.e., the spin-orbit precession torque, which combines the SOT and precession process in an out-of-plane-ferromagnet/spacer/in-plane-ferromagnet composite system [13,27]. These sophisticated theories and technologies have been mostly developed for three-dimensional ferromagnetic metals and oxides. However, it is still largely unknown whether they can be easily expanded to the magnetic vdW materials and their hybrid systems in a compatible way. Moreover, future endeavors shall be made to improve the device's switching performance or simplify the system design, e.g., by removing the in-plane ferromagnet layer. These three appeals are our underlying motivations for conducting the present work.

In this work, we focus on the heterostructure of vdW magnet FGT and oxide surface SrTiO$_3$ (STO) to explore the possibility of its magnetic-field-free switching. The Ar etching can induce a thin conducting layer on top of the STO surface, which hosts sizable spin-charge conversion [28], with the spin Hall angle one order of magnitude larger than that of Pt [29], similar to other 2D conducting STO-based heterointerfaces. In this device, we successfully switch the perpendicular magnetization of FGT without an external magnetic field. We find that the switching polarity of the field-free switching can be reversed when an in-plane magnetic field along or opposite the in-plane current is applied and turned off before the application of the current pulse. This implies that the device has an inherent in-plane interface magnetization, probably near the FGT/STO interface, in addition to the perpendicular magnetization of FGT. The inherent in-plane interface magnetization fulfils the mirror symmetry-breaking requirement for field-free switching. We propose a precession-induced torque mechanism for our system, which combines the inherent in-plane interface magnetization and the Rashba spin-orbit coupling.

***Device fabrication and characterization of the $Fe_3GeTe_2/SrTiO_3$ heterostructure***: We fabricated the field-free switching devices by exfoliating FGT nanoflakes on the Ar-milled STO surface. As depicted in **Figure 1a-d**, first, we performed the Ar-milling treatment on the STO surface by reactive ion etching (RIE) (Figure 1a). The Ar-milling process introduces oxygen vacancies in the STO surface, making the oxide surface conducting (Figure 1B) [30]. Then, FGT nanoflakes were exfoliated on top of the conducting surface by a conventional mechanical exfoliation method with Scotch tape (Figure 1c). Finally, we obtained the FGT/STO heterostructure after peeling off the tape (Figure 1d). Such a vdW magnet/oxide hybrid system is illustrated in a 3D representation in Figure 1e, where the electrodes are patterned by standard e-beam lithography and metal evaporation. We focus on magnetization by monitoring the transverse Hall voltage due to the anomalous Hall effect (AHE) during all transport measurements. Figures 1f and 1g show the optical image and thickness information characterized by atomic force microscopy (AFM) of a typical FGT/STO device. In those devices, the thickness of FGT is roughly around or below 15 nm.

***Magnetic-field-free switching of perpendicular magnetization:*** We performed transport measurements on the fabricated nanodevice to check its ferromagnetic behavior. Figure 1h shows the magnetic field sweeps along the *c*-axis of FGT under an out-of-plane magnetic field. The square-shaped hysteresis loop shows the collective behavior of resistance changes that indicate the perpendicular magnetic anisotropy in the FGT nanoflake. As shown in Figure 1i, the coercive field of the nanodevice decreases with increasing temperature before disappearing at the Curie temperature ($T_c$) of $\sim$ 150 K. Figure 1j shows the remnant resistance $R_{xy}^r$, defined as the Hall resistance, when the magnetic field returns to zero value. When the temperature decreases below the Curie temperature, $R_{xy}^r$ becomes nonzero below $\sim$150 K and sharply rises to the saturated maximum. These observations demonstrate that the nanofabricated device maintains the typical ferromagnetic transport behavior of FGT with perpendicular magnetic anisotropy.

Next, we demonstrate the magnetic-field-free switching of the nanodevice by purely sweeping the writing current using four FGT/STO devices (devices 1, 2, 3, 4). **Figure 2a** illustrates two types of measurement schemes. The upper panel of Figure 2a depicts the magnetic-field-free switching process related to Figure 2b-d: the writing current $I_x$ is applied along the *x* direction, after which the resistance $R_{xy}$ is read out reflecting the change of $m_{\text{FGT}}$. The lower panel of Figure 2a illustrates the anomalous Hall loop shift measurements in Figures 2e and 2f: with a fixed writing current $I_x$, the resistance $R_{xy}$ is monitored while sweeping the external magnetic field $B_{\text{ext}}^z$ along the out-of-plane direction. As can be seen clearly in Figure 2b (device 3), the transverse resistance $R_{xy}$ changes after the applied writing current exceeds 2 mA at 85 K with no magnetic field and reveals a writing-current-dependent hysteresis loop. In the electrical transport measurement, transverse Hall resistance ($R_{xy}$) represents the magnetization of a magnetic device following the AHE mechanism, and thus, Figure 2b unambiguously demonstrates the magnetic field-free switching by current. The critical switching current is defined as the writing current at which $R_{xy}$ meets the middle of the switching.

Figure 2c exhibits the temperature dependence of the critical switching current for devices 1 and 2. The current-induced SOT competes with the magnetic anisotropy of the nanodevice, which decreases as temperature increases. Since the magnetic anisotropy becomes weaker as the temperature increases, the critical switching current reduces upon rising temperature, as shown in Figure 2c. FGT nanoflakes were also exfoliated on the $SiO_2$ substrates instead of the STO substrate. In marked contrast, no current-induced hysteresis loop was observed in FGT/$SiO_2$ samples (Figure 2d, Figure S2 and Figure S3), underlining that the STO substrate, specifically the conducting Ar-milled STO surface, is essential for producing the field-free switching. It is also consistent with previous studies: the intrinsic SOT in FGT does not prefer a particular direction without an external magnetic field [21,23,24]. The above current-driven magnetic-field-free switching indicates that some mechanism in the device is setting a preferred direction. To confirm this point, we measured the magnetic field-dependent hysteresis loop at both positive and negative currents supplied by a D.C. current source.

Figure 2e displays the hysteresis loops when the charge currents of +0.01 mA and -0.01 mA are injected into the device (device 1). The hysteresis loops of the two cases are very similar, manifesting that the current-induced spin polarization is too weak to shift the AHE loops under small currents below the critical switching current. In Figure 2f, we apply charge currents of +2 and -2 mA to the same device in Figure 2e while sweeping the magnetic field. In sharp contrast, positive and negative currents shift the AHE loop oppositely. The raw data of Figures 2e and 2f is presented in Figure S4 of the Supporting Information, showing the same behavior. Similar phenomena were reproduced in Figure S5 for another device (device 3). Such a current direction-governed shift of the AHE loop proves the presence of a preferred direction induced by an in-plane current injection of STO conducting layers. When we apply a positive current, the loop shifts to the left side, showing the preference toward the upward direction.

Conversely, the opposite current shifts the loop to the right side, indicating the preference toward the downward direction. These observations on the preferred direction agree with switching polarities in the magnetic-field-free switching measurements in Figures 2b and S1. We note that it is unclear whether the preferred direction is determined by the field-like torque with the effective out-of-plane field or the damping-like torque with the out-of-plane spin polarization since both can induce the shift of the AHE loop. All these observations are consistent with the prototypical SOT field-free switching, excluding possible experimental artefacts in our FGT/STO system, as detailed in Supporting Note 1.

***Switching polarity dependence of the in-plane magnetic field initialization:*** As described above, we have demonstrated the current-induced anomalous Hall loop shift in Figure 2f. The Ar-milled STO surface exhibits strong Rashba spin-orbit coupling [28]. When a current is injected in the $x$ direction, spins aligned with the $y$ direction accumulate at the interface due to its spin-orbit coupling. However, such spin-polarized accumulation along the $y$ direction cannot directly generate an out-of-plane SOT necessary for field-free switching. Additional symmetry must be broken to achieve field-free switching, e.g., by introducing additional magnetism. This additional symmetry breaking produces a preference for magnetization direction in the FM layer depending on the charge current direction. Next, we show that this additional symmetry breaking can arise from the emergence of in-plane interfacial magnetism $m_{\text{interface}}$ between FGT and STO, and the switching polarity can be controlled by the in-plane magnetic field initialization.

**Figure 3a** illustrates the measurement schematic of switching polarity's dependence on in-plane magnetic field initialization. To put it simply, the in-plane magnetic field $B_{ext}$ is applied along the $-x$ direction, forcing the $m_{interface}$ along the $-x$ direction; then, the magnetic field is turned off while $m_{interface}$ remains along the $-x$ direction. After this in-plane magnetic field initialization, we performed the magnetic-field-free switching, as shown in Figure 2b. Similarly, the in-plane initializing magnetic field $B_{ext}$ is applied along the $x$ direction, eventually making the $m_{interface}$ along the $x$ direction after the in-plane magnetic field initialization.

With $m_{interface}$, we depict the possible mechanism for magnetic-field-free switching in Figure 3b. Firstly, charge current $J_e$ generates well-defined spin accumulation, $\sigma_{Rashba}$, by Rashba spin-orbit couplings at the STO surface. Secondly, the accumulated spins $\sigma_{Rashba}$ can diffuse in the $z$ direction indicated by the spin current $J_s$. During diffusion to the top ferromagnetic layer, the spin experiences a torque by interface magnetization aligned $x$-direction. The direction of the torques, $\tau_m = m_{interface} \times \sigma_{Rashba}$, varies depending on the directions of both the in-plane current injection $J_e$ and the magnetization at the interface $m_{interface}$. Specifically, the torques reverse their directions whenever either the current injection direction or $m_{interface}$ is reversed. This variation in the torque direction, dependent on both directions, determines the out-of-plane spin polarization directions of the spins injected into the ferromagnetic layer. Finally, the injected spins with the out-of-plane polarization exert a damping-like torque on the top ferromagnetic layer, eventually making the magnetic field-free switching. In short, the $m_{interface}$'s direction will regulate the switching polarity if the above mechanism works for our case.

**Figure 4a** shows the dependence of switching polarity on the $m_{interface}$'s orientation initialized by an external magnetic field for device 1 (See more details with raw data in Figure

S8). First, we applied a magnetic field of +8 kOe to initialize the configuration and then turned off the field to conduct the current sweep measurement with zero magnetic field. The switching polarity after the +8 kOe initialization is counterclockwise, in contrast to the clockwise switching polarity after the -8 kOe initialization. We see some small random peaks near the switching current, whose origins have been discussed in Supporting Note 2. Such switching polarity change behavior was reproduced in another device (device 4) with raw data shown in Figure S9. By repeating the above measurements with varying the in-plane initializing magnetic fields $B_{ext}$, we can obtain the switching polarity as a function of $B_{ext}$ in Figure 4b (see more details in all the switching curves and raw data in Figure S10) for device 4. We emphasize that the field-free switching by current and the switching polarity change by an in-plane magnetic field are reproduced by several experiments in our work. Interestingly, the switching polarity shows a hysteresis loop regarding $B_{ext}$ and only changes when $B_{ext}$ is larger than the coercivity of ~30 Oe. The observation of the anomalous Hall loop shift in Figure 2f and the coercivity of switching polarity in Figure 4b are consistent with the switching mechanism described in Figure 3b. These findings also agree with the previous reports of the perpendicular switching by the external in-plane magnetic field or ferromagnet heterostructure [7,27]. The FGT/STO device indeed hosts both the out-of-plane polarization and the hysteresis of polarization switching, suggesting the existence of in-plane magnetization $m_{interface}$ between FGT/STO. We have discussed the in-plane interface magnetism issue in Supporting Note 3. We have also discussed the Joule heating effect in our field-free switching experiments in Supporting Note 4.

**Discussion:** As shown in Figure 3b, the torques, which induce precession of spin, possess an opposite direction when the in-plane magnetization ($m_{interface}$) is flipped. Among physical quantities, magnetization reacts to an external magnetic field and manifests coercivity against the magnetic field. Therefore, the polarization changes with a larger magnetic field than the coercivity, indicating interfacial magnetization in an in-plane direction. The possible origin can come from interfacial interaction, charge transfer, or oxide vacancy, similar to the heterostructure effect reported in the LAO/STO structure [31,32]. Our work unmistakably demonstrates the magnetic-field-free switching enabled by the current-induced out-of-plane spin polarization mediated by the $m_{interface}$-dependent SOT. Specifically, the current-induced spin polarization by Rashba spin-orbit coupling can be changed to the out-of-plane direction suitable for perpendicular switching via precession-facilitated SOT. Moreover, our lowest critical switching current density was estimated to be $\sim 6 \times 10^{10}$ A/m$^2$, which is significantly lower than previous switching studies [27,33]. The reduction of critical switching current can benefit from the large spin Hall angle in STO, which is why we initially used STO. There could also be contributions from intrinsic SOT in FGT [21,23,24] and domain nucleation [5]. Additionally, the Joule heating may also help lower the switching current by reducing the free energy barrier of FGT. Although a variety of research for large spin Hall angle materials has been conducted, most materials only possess ordinary spin polarization. Our work suggests the applicability of previously reported perovskite materials with large spin Hall angles to achieve magnetic-field-free and energy-efficient switching.

In summary, we have demonstrated magnetic field-free perpendicular magnetization switching of a heterostructure device composed of FGT and Ar-milled STO surface. It is the first vdW magnet and oxide materials combined directly for the SOT. Our novel approach unitizes the large Rashba spin-orbit coupling of STO and high energy efficiency in FGT to realize energy-efficient field-free switching. The spintronics concept can be successfully employed for this novel hybridized system. Our work will prompt an interesting new direction of active research in new material combinations for magnetic-field-free SOT switching.

**Experimental Section**

*Sample Preparation:* FGT crystals were synthesized using the chemical vapour transport method with iodine as a transport agent [34,35]. High-purity powder elements, Fe (99.9%, Thermo Scientific), Ge (99.999%, Thermo Scientific), and Te (99.999%, Thermo Scientific), were prepared according to the stoichiometric mass ratio and sealed in a glove box with Ar condition (smaller than 1 ppm for $H_2O$ and $O_2$). Sealed powders were heated with a temperature gradient from 700 ºC to 750 ºC for 7 days in a two-zone furnace.

SrTiO$_3$ (001) crystals were purchased from Crystal GmbH. SrTiO$_3$(001) surface was exposed to Ar plasma using an RIE machine under 200 mTorr pressure, 50 W power, and a 10 s duration. Using the Scotch tape method, FGT crystal was exfoliated on the Ar plasma-treated surface of SrTiO$_3$(001). EBL (Electron Beam Lithography) was conducted after coating PMMA A7 and E-spacer to pattern the electrodes. Electrodes were deposited with Ti/Au(5/10nm) using an E-beam evaporator.

*Electrical Transport measurement:* The transport measurements were performed using a resistance probe placed inside a closed circuit refrigerator (CCR), which can lower the temperature to 3 K. The temperature of the CCR was regulated by a LakeShore 340 Temperature Controller by PID control. The magnetic field was applied by an electromagnet and monitored by Lakeshore 450 Gaussmeter. The Lock-in amplifier SR830 and D.C. Current Source Keithley 6220 were used for monitoring $R_{xy}$ and applying D.C. current for experiments mentioned in Figures 2b, 2d, 4a, and 4b. In these figures, we turned off the writing current and then waited 20 seconds before recording $R_{xy}$. Then, $R_{xy}$ was recorded as the average value over 20 seconds using a 10 μA reading current. The nanovoltmeter Keithley 2182A, and D.C. current source Keithley 6220 were used for monitoring $R_{xy}$ and applying D.C. current in Figures 2e and 2f.

**Supporting Information**
Supporting Information is available from the Wiley Online Library or the author.

**Acknowledgments**
J.K. and K.Z. contributed equally to this work. We acknowledge Jaehoon Park for the helpful discussion. The work at CQM and SNU was supported by the Samsung Science & Technology Foundation (Grant No. SSTF-BA2101-05). One of the authors (J.-G.P.) was partly funded by the Leading Researcher Program of the National Research Foundation of Korea (Grant No. 2020R1A3B2079375). The theoretical works at the POSTECH were funded by the National Research Foundation (NRF) of Korea (Grant No. 2020R1A2C2013484). In addition, the Samsung Advanced Institute of Technology also supported this work at both SNU and POSTECH.

Received: ((will be filled in by the editorial staff))
Revised: ((will be filled in by the editorial staff))
Published online: ((will be filled in by the editorial staff))

**Figures**

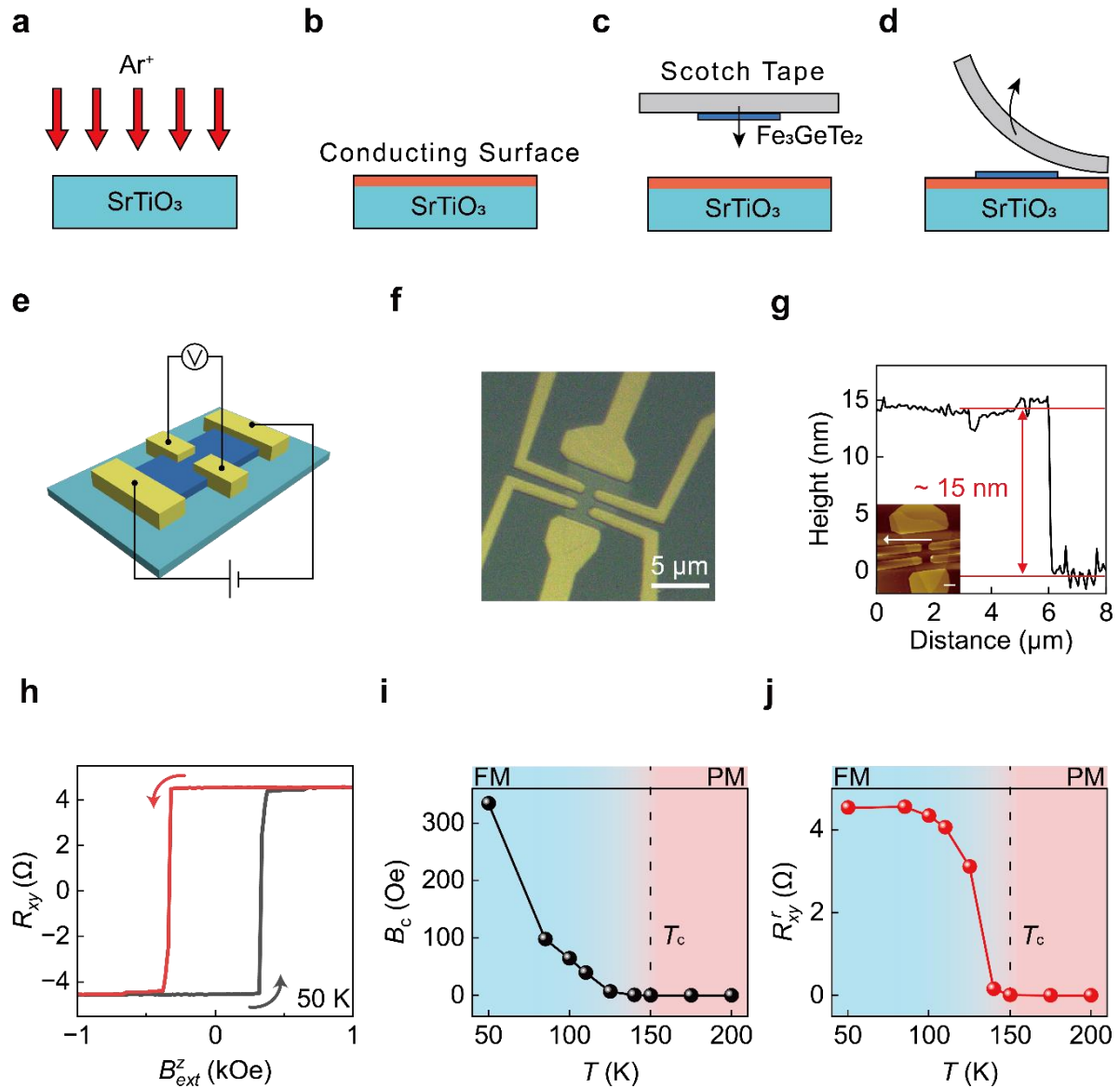

**Figure 1.** Fabrication process for the FGT/SrTiO₃(001) heterostructure and its prototypical ferromagnetic behaviors. a-d) A two-dimensional surface conducting layer was formed by Reactive ion etching with Ar plasma on SrTiO₃(001) surface (a, b). A thin layer of FGT was exfoliated on top of SrTiO₃ using the Scotch tape method (c, d). e) Schematic of the field-free switching device. f) Optical microscope image of a typical device with Hall electrodes. g) Atomic force microscopy imaging of the corresponding device in (f), where the thickness was measured to be about 15 nm. h) Anomalous Hall loop ($R_{xy}$-$B_{\text{ext}}^z$ curve) of the device. A magnetic field is applied along the c-axis of FGT. The device shows a square loop and perpendicular anisotropy at 50 K. i) Temperature dependence of coercivity $B_c$ for the device. In the figure, FM refers to the ferromagnetic phase, and PM refers to the paramagnetic phase. The coercivity disappears around 150 K. j) Temperature dependence of remnant Hall resistance ($R_{xy}^r$), which vanishes above around 150 K. The transport properties of devices show the similar behavior to those of typical FGTs.

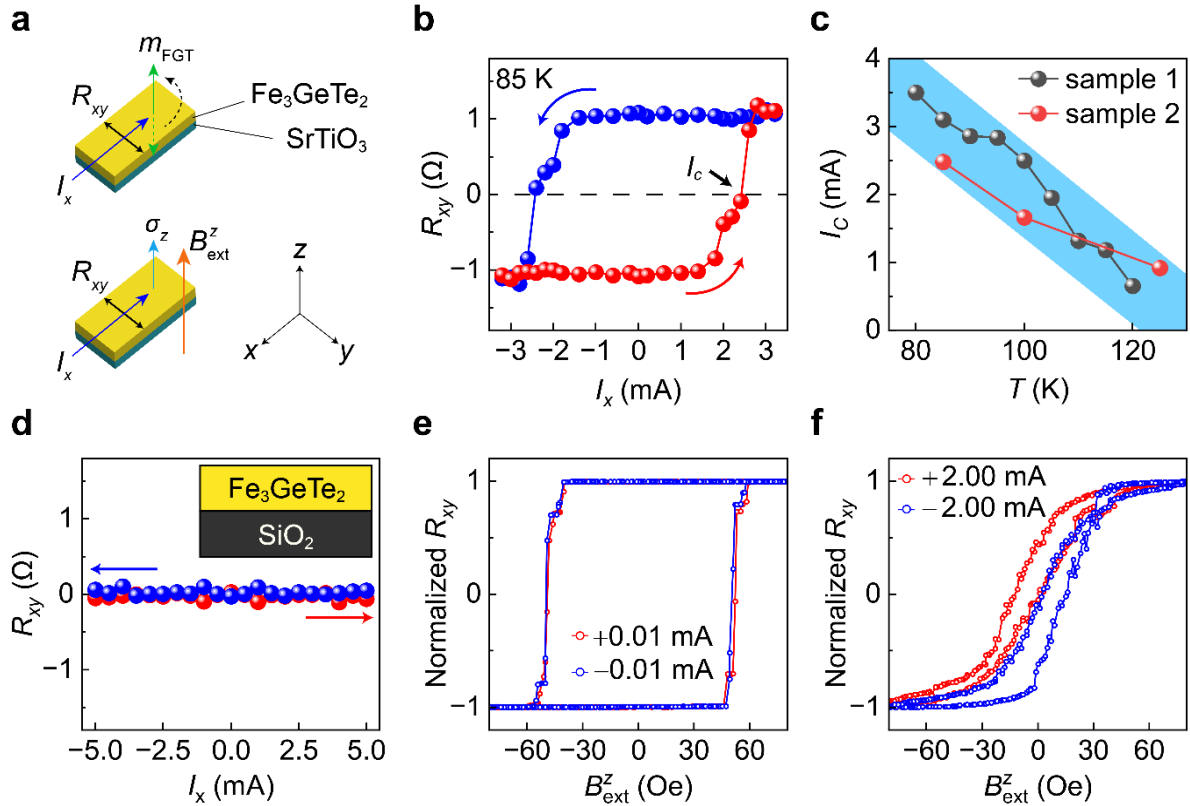

**Figure 2**. Field-free switching demonstration of the FGT/SrTiO₃ device and anomalous Hall loop shifts by current-induced out-of-plane spin polarization. a) Schematics of magnetic-field-free switching and out-of-plane anomalous Hall loop shift measurement. The above illustration indicates that in-plane current injection (blue arrow) controls perpendicular magnetization (green) of FGT. The change in magnetization $m_{FGT}$ can be monitored by $R_{xy}$ (black). The illustration below depicts anomalous Hall loop shift measurement. Out of the plane directional spin polarization, $\sigma_z$ (light blue), is induced by in-plane current. This spin polarization creates asymmetry in the coercive field. b) Antisymmetrized Hall resistance $R_{xy}$ of the device 1 as a function of writing current $I_x$ without magnetic field at 85 K for device 3. $R_{xy}$ or magnetization is switched by the current direction, and the critical switching current is defined as $I_c$, at which the $R_{xy}$ changes to the middle value. c) $I_c$ versus temperature for devices 1 and 2. The switching current decreases as temperature increases, consistent with the decreasing coercivity upon increasing temperature. d) Hall resistance $R_{xy}$ as a function of writing current

measured in another sample, i.e., the FGT/bare-SiO$_2$ substrate. $R_{xy}$ shows a nearly flat background regardless of a current sweep, demonstrating no field-free switching by current. e) Anomalous Hall loops at 110 K under a D.C. current of +0.01 mA (red) and -0.01 mA (blue) for device 1. The two loops are mutually overlapped. f) Anomalous Hall loops of the device 2 at 110 K under a D.C. current of 2.00 mA (red) and -2.00 mA (blue) for device 1. Positive current pushes the loops to the left while negative current shifts the loop opposite. The shift between the blue and red loops is 14.6 Oe.

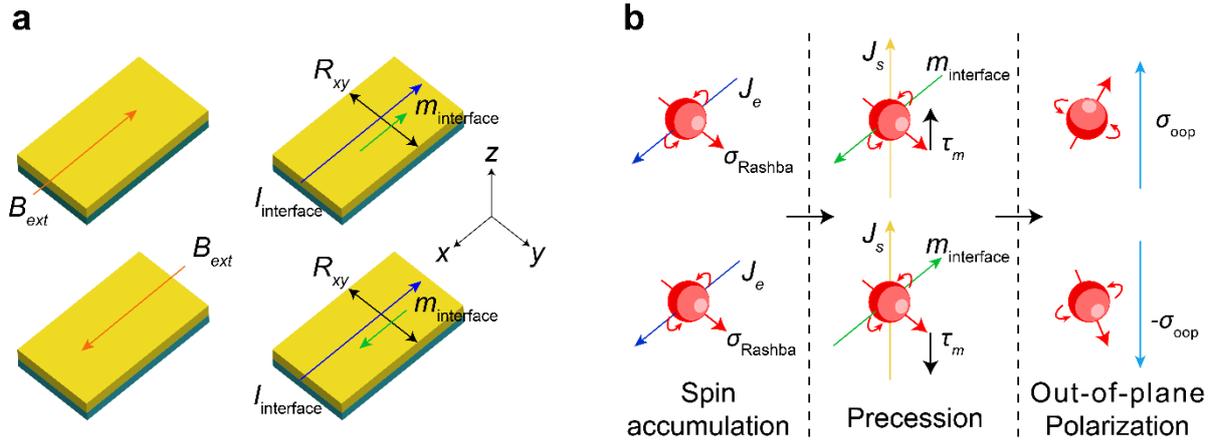

**Figure 3.** Switching polarity's dependence on the interfacial magnetism controlled by in-plane magnetic field initialization. a) Schematics of measurement. An external magnetic field is applied in both $x$ and -$x$ directions. After turning off the external magnetic field, a charge current is injected along the $xy$ plane to observe switching polarity. b) Schematics of switching mechanism. Charge current, $J_e$, induces spin accumulation by Rashba spin-orbit coupling. After that, the interfacial magnetization, $m_{\text{interface}}$, exerts a torque $\tau_m = m_{\text{interface}} \times \sigma_{\text{Rashba}}$ on the accumulated spin. The spins polarized by Rashba spin-orbit coupling undergo precession by the torque, $\tau_m$. Finally, the precession of spins creates an out-of-plane spin polarization.

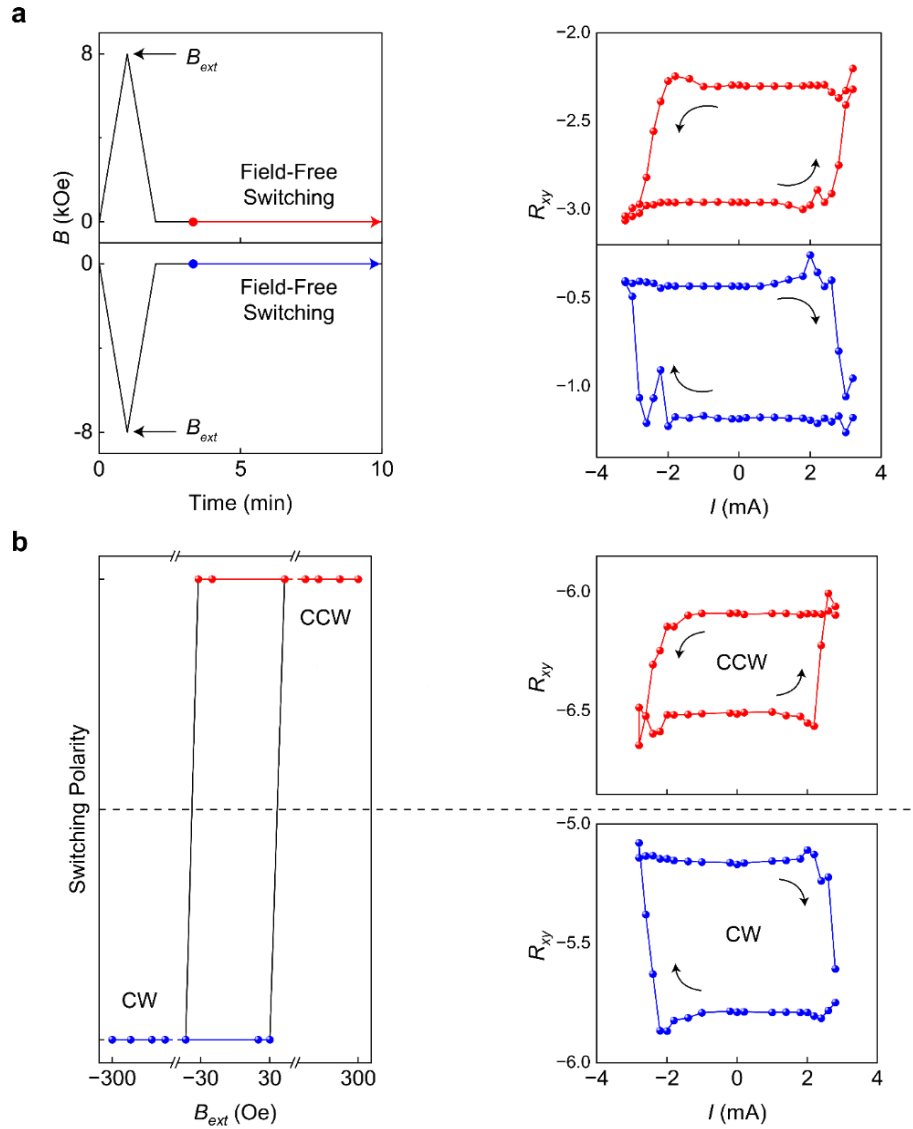

**Figure 4.** Experimental results of the switching polarity change. a) Left panel for measurement schematic: First, we initialize the system by applying an in-plane magnetic field of a $0 \rightarrow B_{ext}$ $\rightarrow 0$ sequence. Second, we sweep the writing current under a zero magnetic field to measure field-free switching. The black arrows indicate the $B_{ext}$, i.e., the peak value of the applied magnetic field of 8 kOe. Circles indicate the starting point of current sweep measurements. Right panel for measurement results: The switching polarity of the device 1 at 90 K was observed to be CCW (counterclockwise) and CW (clockwise), respectively. Switching polarities vary depending on the initialization magnetic field direction. b) Hysteresis of switching polarities for the device 4 at 120 K, shown in the left panel. In the hysteresis loop,

the positive region is defined as CW, and the negative region as CCW. The switching polarity coercivity was measured to be 30 Oe, indicating that it has certain magnetic anisotropy. The right panels represent CCW and CW switching behavior.

**The table of contents entry**

In-plane current-induced perpendicular magnetization switching under zero magnetic field is highly desired for industrial applications. For the first time, field-free switching is realized in an unprecedented combination of a van-der-Waals/oxide heterostructure. A comprehensive investigation reveals the underlying mechanism, which is precession-induced out-of-plane spin polarization and consequent unconventional spin-orbit torque. Our work demonstrates a novel approach to utilize both the interface and the material with large Rashba spin-orbit coupling for energy-efficient field-free switching. It opens a new research realm by combining burgeoning van-der-Waals magnets and superior oxide spintronics.

Jihoon Keum, Kai-Xuan Zhang,* Suik Cheon, Hyunchoel Kim, Jingyuan Cui, Giung Park, Yunyeong Chang, Miyoung Kim, Hyun-Woo Lee and Je-Geun Park*

Title: Novel magnetic-field-free switching behavior in vdW-magnet/oxide heterostructure

ToC figure

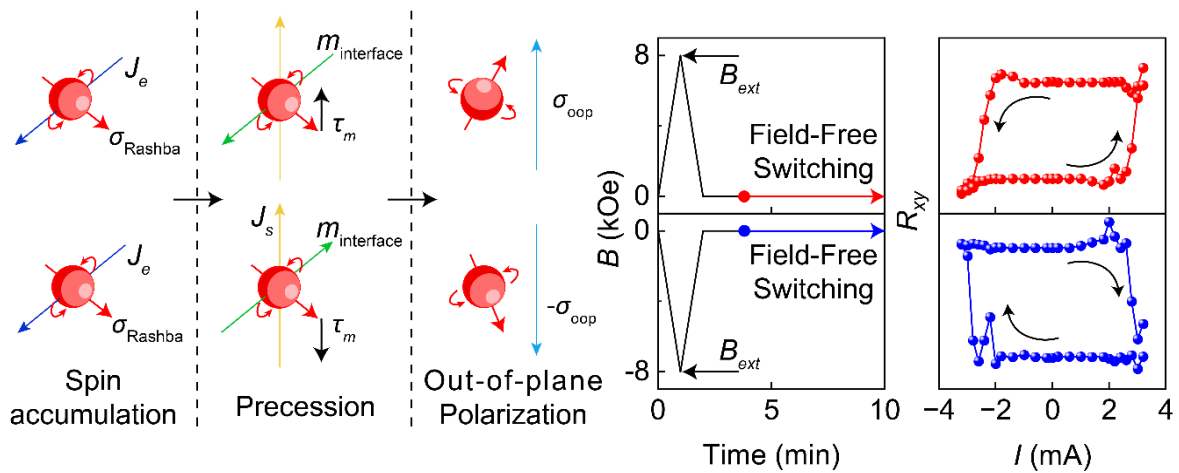

# Supporting Information

**Novel magnetic-field-free switching behavior in vdW-magnet/oxide heterostructure**

*Jihoon Keum, Kai-Xuan Zhang,\* Suik Cheon, Hyuncheol Kim, Jingyuan Cui, Giung Park, Yunyeong Chang, Miyoung Kim, Hyun-Woo Lee and Je-Geun Park\**

J.K. and K.Z. contributed equally to this work.

This file includes:

1. Supporting Notes 1-4

2. Supporting Figures S1- S16

3. Supporting References

**Supporting Notes**

**Note 1. Experimental evidences/facts for validating the field-free switching.**

We summarise several of the most important experimental evidence and facts that strengthen and validate our field-free switching contents as follows:

(1) We have shown that the Hall resistance can be changed by pure current without a magnetic field, and the change's direction is determined by the current polarisation [Figure 2b].

(2) We have demonstrated that current can produce the shift of anomalous Hall loop [Figure 2e and 2f]. Meanwhile, the shift direction is regulated by the current polarisation, which exhibits consistent dependence on the current polarisation of the Hall resistance change in point (1). Generally, both points (1) and (2) are considered sufficient evidence for the unambiguous demonstration of intrinsic field-free switching: please refer to the following previous studies on field-free switching[1,2].

(3) Additionally, we provided a comparison device for the FGT/silica substrate instead of the FGT/Ar-STO substrate. In such a comparison device, pure current does not produce field-free switching [Figure 2d], which excludes the measurement artefacts in our studies with FGT/Ar-STO systems.

(4) Moreover, no field-free switching is observed above the Curie temperature $T_c$ [Figure S14], demonstrating its intimate relationship with magnetism and excluding the experimental artefacts once again.

(5) Furthermore, the switching current increases with lowering temperature [Figure 2c], which is consistent with the fact that the coercivity or magnetic anisotropy or switching barrier enhances at lower temperatures, requiring larger switching currents. This observation excludes the experimental artefacts a third time.

In summary, all these observations constitute extremely strong evidence in favor of our conclusion and point to a field-free switching phenomenon without doubt for the FGT/Ar-STO system.

**Note 2. Possible trivial origins of the random small peaks near the switching current.**

During the field-free switching, the Joule heating induced by the applied current can create multidomain or demagnetisation in FGT. Additionally, either the pinning effect caused by unaltered neighboring domains or random domain fluctuation/motion in the Hall bar region can cause certain resistance changes in the Hall bar region. These phenomena and accidental domain fluctuation can cause unexpected resistance fluctuation, especially the small sharp peaks at the large critical switching current. However, we emphasise that the clear dominant $R_{xy}$ change, or the magnetisation switching, is deterministically controlled by the writing current polarisation, which is clearly distinguished from the random effect discussed above. Similar fluctuations also occur in previous studies using van der Waals ferromagnet[3].

**Note 3. Discussions on the in-plane interface magnetism.**

Here we want to discuss the in-plane interface magnetism issue as follows:

(1) To investigate the origin of the interface magnetisation, it is better to employ the magnetic element, specifically X-ray magnetic circular dichroism (XMCD). However, it is very challenging to obtain magnetic signals for our FGT/STO devices using currently available techniques because the oxide is covered with metallic FGT nanoflakes thicker than 10 nm and narrower than 10 μm. This prevents light penetration into the interface and decreases signal amplitude from a small beam spot.

(2) Although we couldn't directly detect the in-plane interface magnetisation by using the XMCD method due to technical challenges, we alternatively provide the transport evidence of such magnetism via the switching polarity change driven by in-plane field initialisation in Figure 4a and the hysteresis of such effect in Figure 4b. According to Figure S7 and Figure S9, the relatively small in-plane magnetic field does not change FGT's magnetisation much. Thus, its net effect is to alter the interface magnetism for switching polarity change experiments in Fig. 4. Therefore, switching polarity change by in-plane magnetic field initialisation is a useful transport probe for detecting the interface magnetism in our FGT/STO devices. We believe that these results provide reasonable evidence for the existence of in-plane interface magnetisation. In addition, Figures 3e and S5 show a certain exchange bias effect, although very weak: the difference in the coercivity under positive and negative magnetic fields was approximately 20 Oe for our devices. Such a small exchange bias effect can provide additional evidence supporting the existence of interface magnetism.

(3) Moreover, we have performed the field-free switching measurement of FGT/STO without an Ar-milled surface. In this case, no field-free switching is observed (Figure S11), which supports our claim that the conducting (Ar-milled) STO surface interfacing with FGT is important for field-free switching. This observation also points to the possibility that Ar-etching and interfacing with FGT produces the in-plane interface magnetisation.

(4) Furthermore, we have performed the cross-sectional scanning transmission electron microscopy (STEM) characterisation of the FGT/STO interface (Figure S13). According to TEM images, we can observe an amorphous interface created by Ar milling between STO and FGT. The EDS results show that the ratio of Sr and Ti at the interface created by Ar milling is reduced compared to areas where the $SrTiO_3$ crystal planes are clearly visible. We think such a reduction in Sr and Ti elements can induce magnetism[4], suggesting that the origin of the in-plane magnetisation most likely comes from Ar-milled $SrTiO_3$ surface defects interfacing with FGT. The interface magnetism is sometimes fragile but sometimes can exist at high temperatures. For example, a FGT/$Bi_2Te_3$ heterostructure can raise $T_c$ from below 200 K to 300 K[5]. Still, this problem is an interesting future direction, which unfortunately falls out of our present capacities.

**Note 4. Discussions on the Joule heating issue.**

We believe that the Joule heating caused by the current does not significantly affect our main results and conclusions for the following reasons:

(1) The Joule heating effect is unavoidable in current-related SOT spintronics and has been heavily discussed in many references[6-8]. As naturally expected, and also shown in the references, the Joule heating exhibits symmetric behavior regardless of current polarisation. Still, we observed an anti-symmetric switching behavior in our work: positive/negative current tends to switch magnetisation toward the up/down direction.

(2) The switching current of around 1-3 mA in our work is within a reasonable range compared to previous field-free SOT switching work of FGT with around 5 mA[2]. Therefore, our work is consistent with the previous classic works regarding the Joule heating issue.

(3) Finally, we have another strong evidence in our own data: The anomalous Hall loop shift was measured under 3.6 mA (Device 3) [Figure S5], larger than the switching current of 1.5 mA [Figure S1c] for Device 3 at 120 K. Since the anomalous Hall loop exists even with a larger current of 3.6 mA, during the field-free magnetisation switching by a smaller current of 1.5 mA, the temperature raised by the Joule heating is still below the $T_c$ and the system maintains in the ferromagnetic phase. This observation provides unambiguous evidence that our field-free switching is not coming from the Joule heating.

**Supporting Figures**

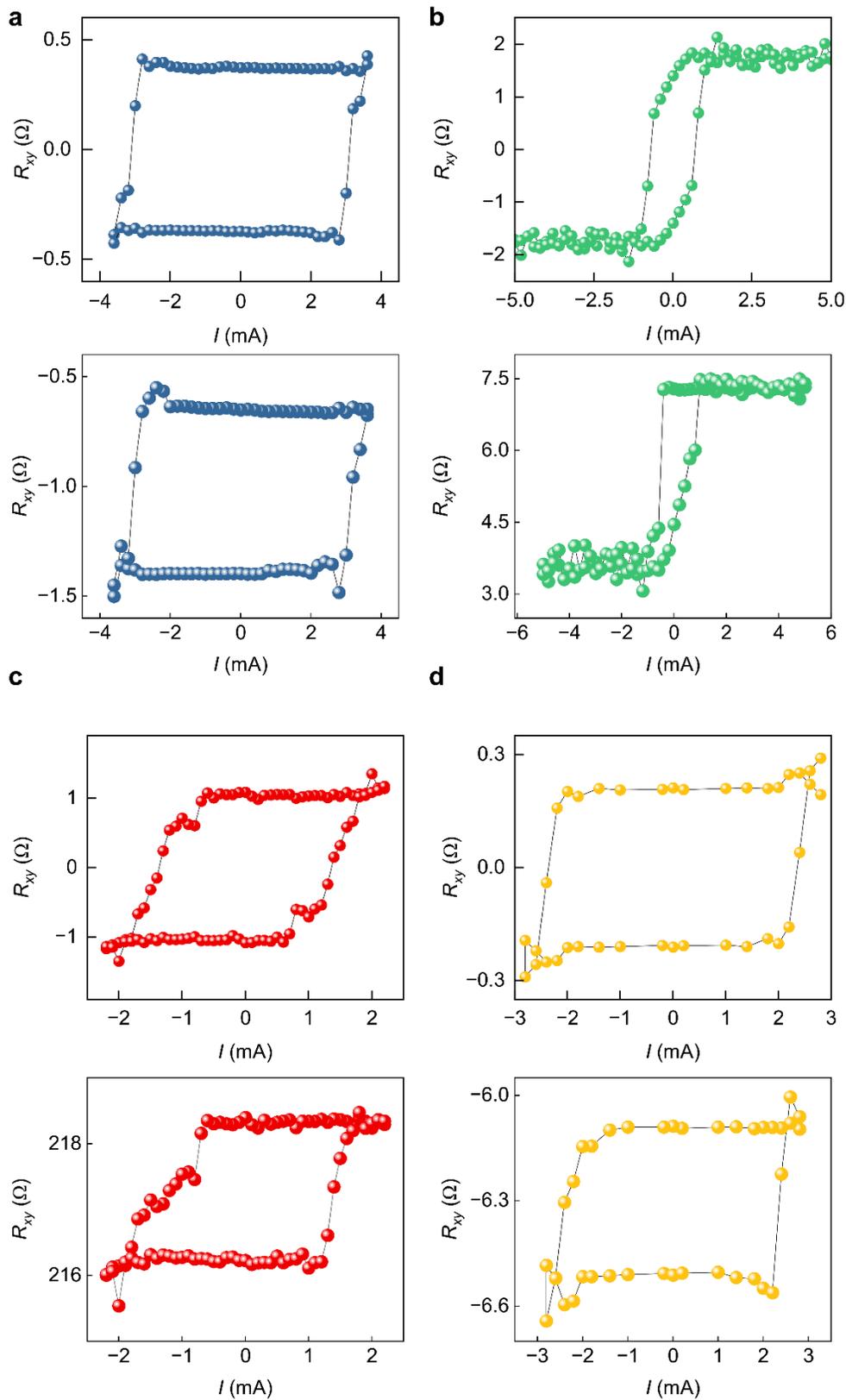

**Figure S1.** Field-free switching behaviors for Devices 1, 2, 3, and 4. a-d) Antisymmetrized Hall Resistance $R_{xy}$ (upper panels) and Raw data (lower panels) as a function of writing current $I$, without external magnetic field for Device 1 at 85 K (a), Device 2 at 125 K (b), Device 3 at 120 K (c), and Device 4 at 120 K (d).

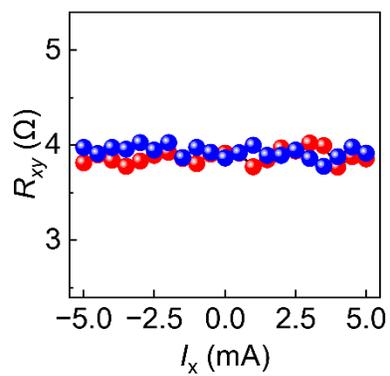

**Figure S2.** $R_{xy}$ raw data for sample using in Figure 2d as a function of writing current $I$, without external magnetic field.

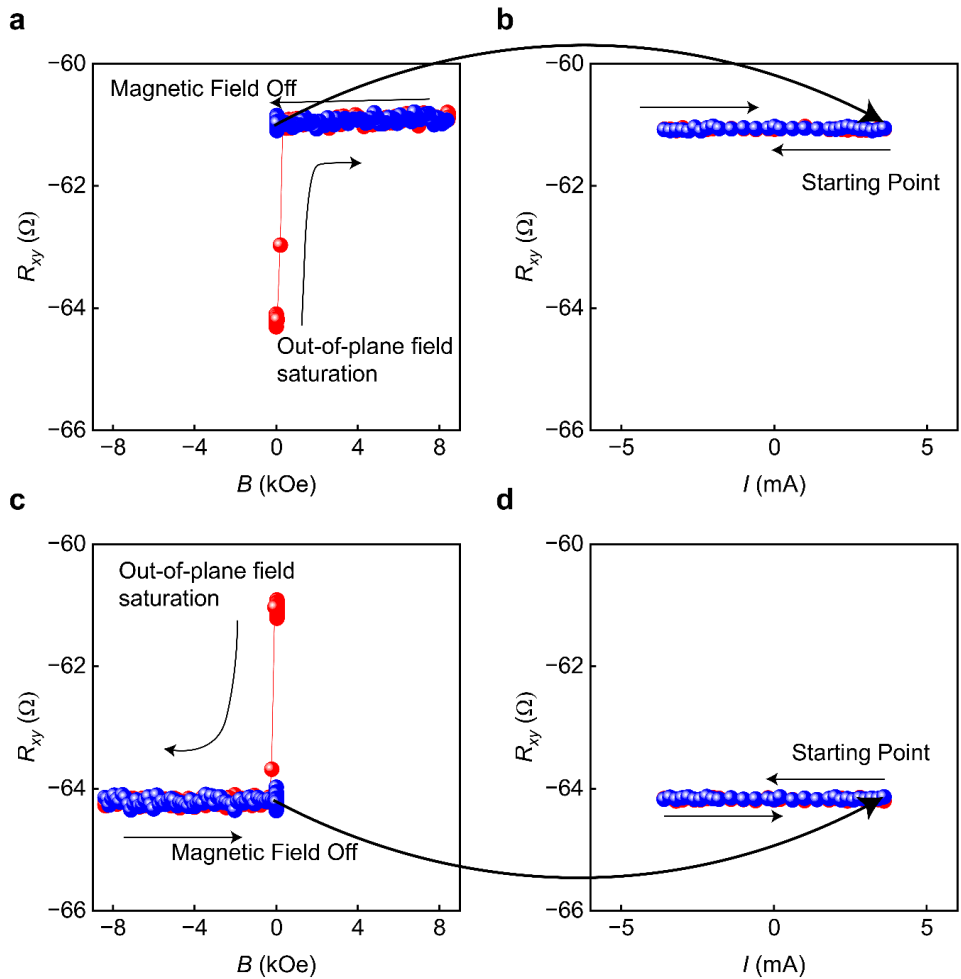

**Figure S3.** Magnetisation saturation by the out-of-plane magnetic field and the field-free switching measurement of the FGT/SiO$_2$ comparison device at 85 K. a) -8 kOe magnetic field was applied and then turned off. b) Starting from 3.6 mA, the current decreased in steps of 0.2 mA to -3.6 mA and then increased back to 3.6 mA. c) Afterwards, 8 kOe magnetic field was applied and turned off. d) Starting from 3.6 mA, a current sweep was performed.

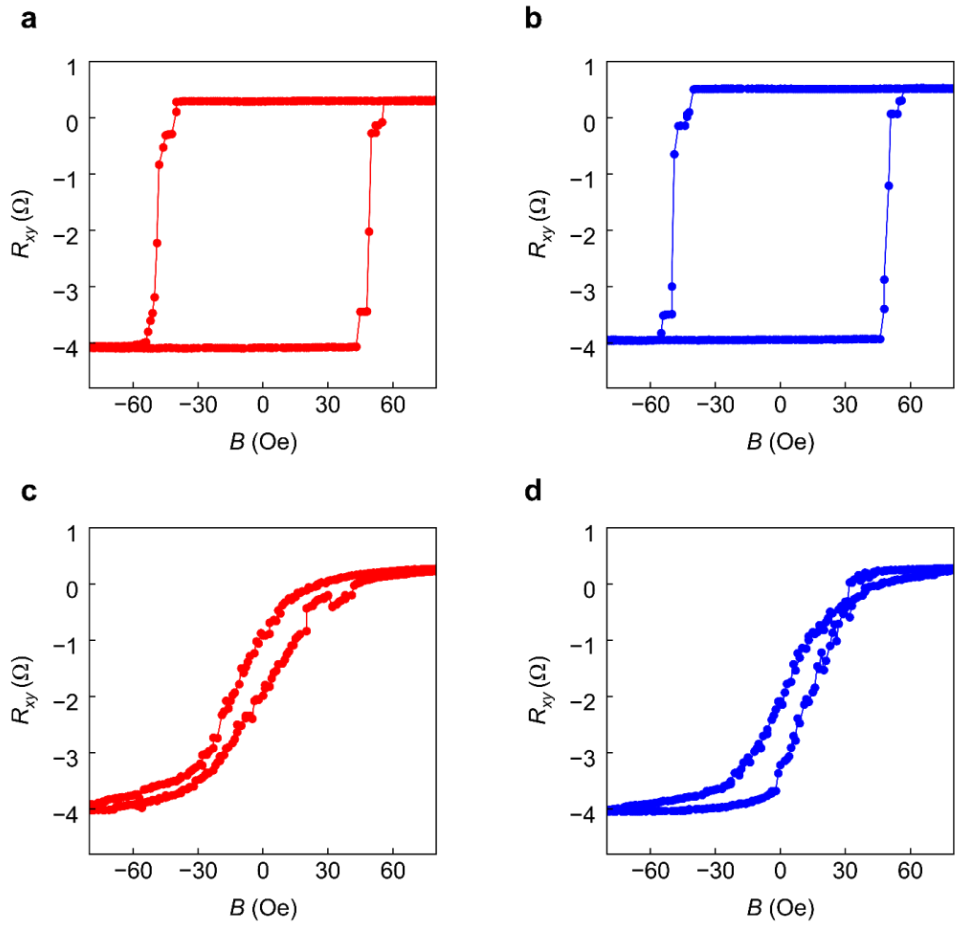

**Figure S4.** Raw data of Figure 2e and 2f. a-b) Anomalous Hall loops of Device 1 at 110 K under a DC current of +0.01 mA (a) and -0.01 mA (b). c-d) Anomalous Hall loops of Device 1 at 110 K under a DC current of +2 mA (c) and -2 mA (d).

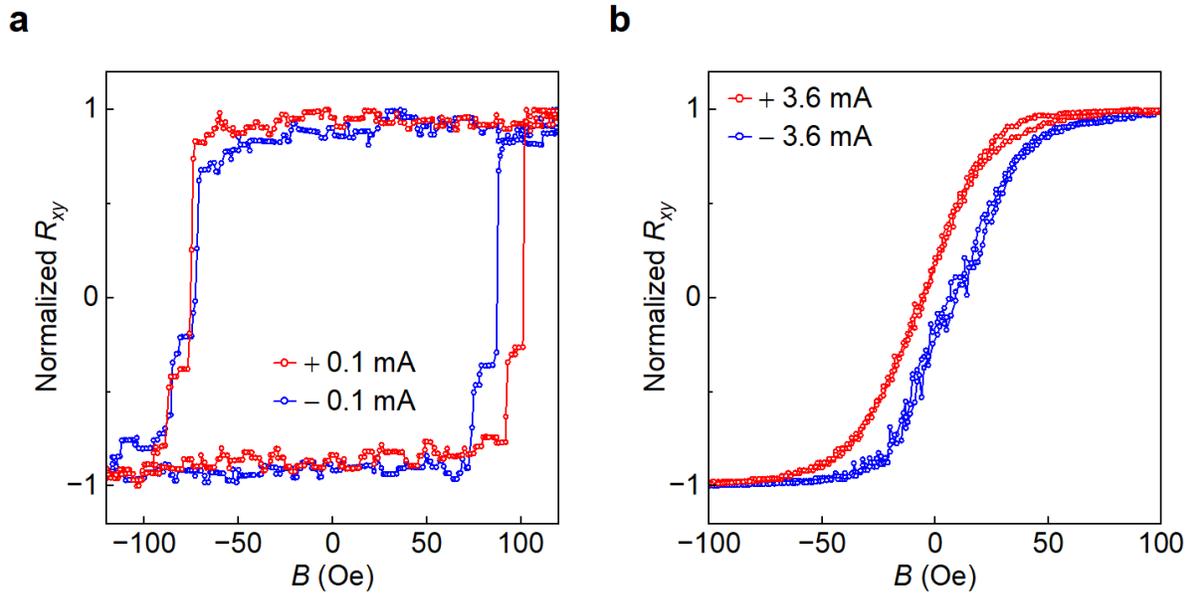

**Figure S5.** Anomalous Hall loop shifts by current-induced out-of-plane spin polarisation of Device 3. a-b) Anomalous Hall Loops at 120 K under a DC current of 0.1 mA (red) and -0.1 mA (blue) in (a), and 3.6 mA(red) and -3.6 mA (blue) in (b). It demonstrates the reproducibility of the anomalous Hall loop shifts.

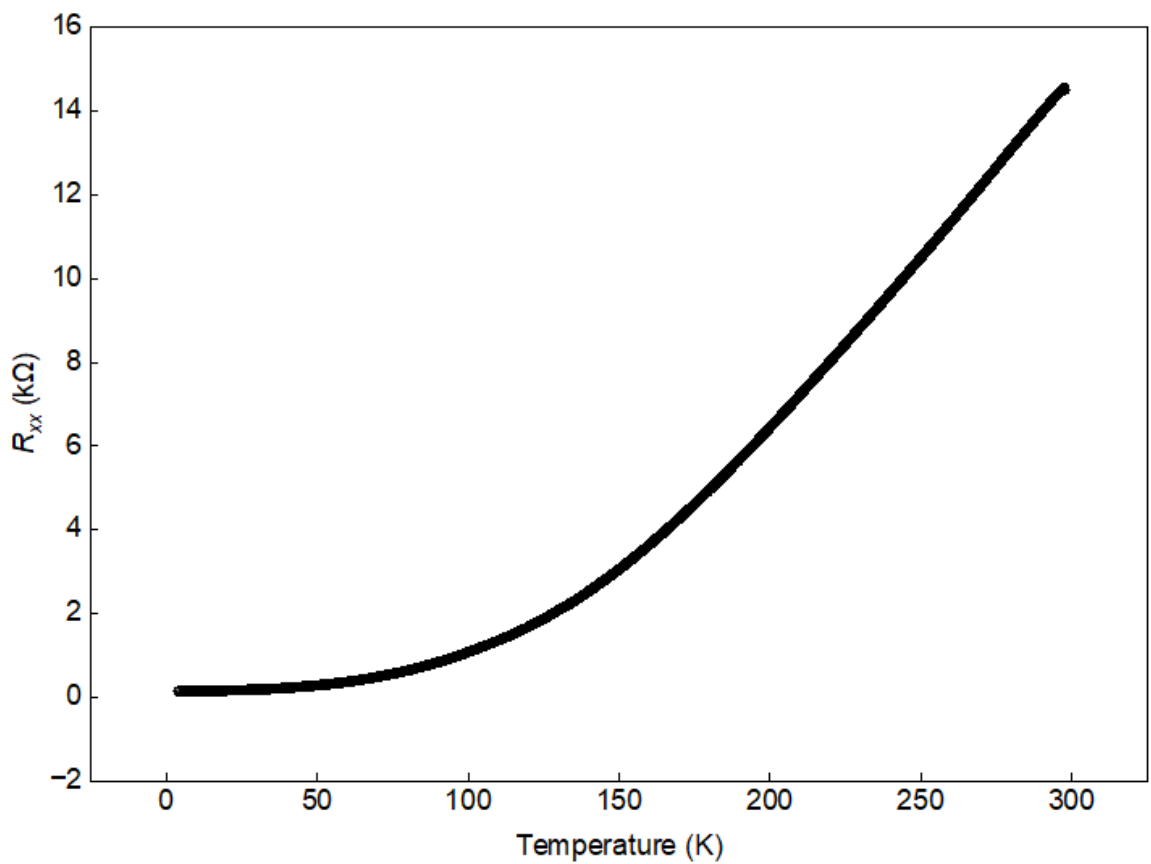

**Figure S6.** $R_{xx}$-$T$ curve in Ar-etched SrTiO$_3$. It shows a metallic behavior in longitudinal

resistance $R_{xx}$. The sample geometry has a width of 2.8 mm, and the distance between probes is

2.1 mm.

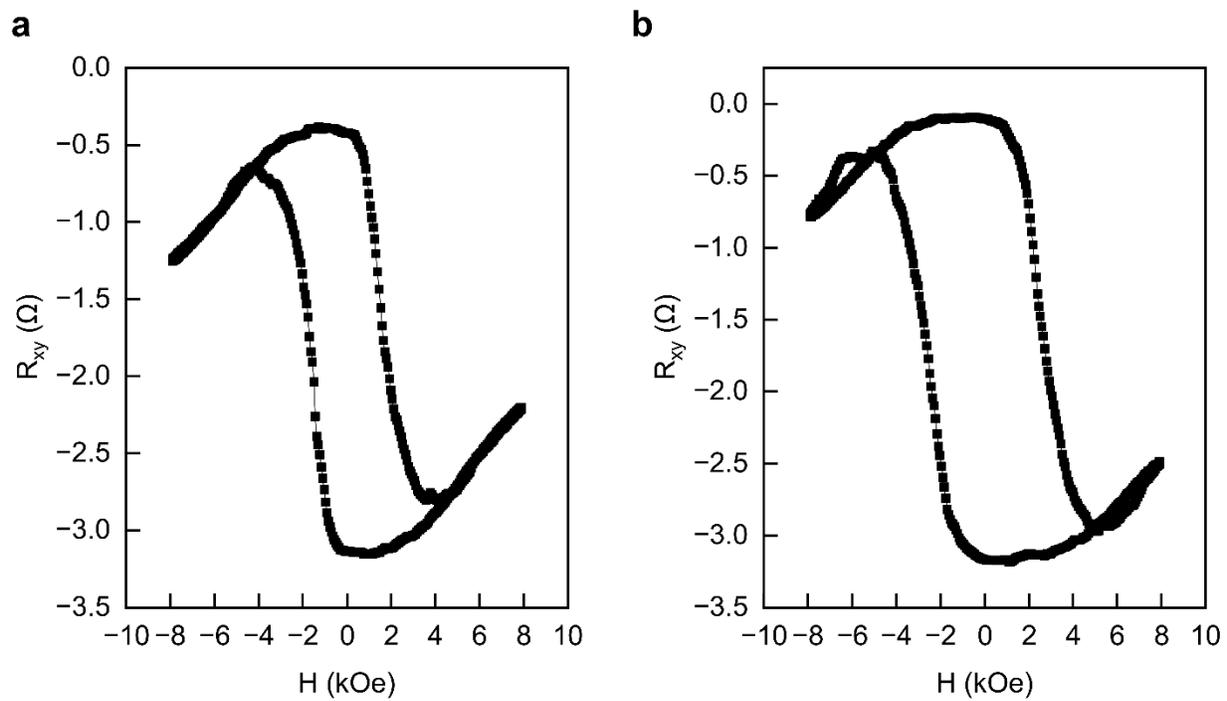

**Figure S7.** $R_{xy}$ under in-plane magnetic field sweeps in Device 1 at 110 K (a), 90 K (b).

WILEY-VCH

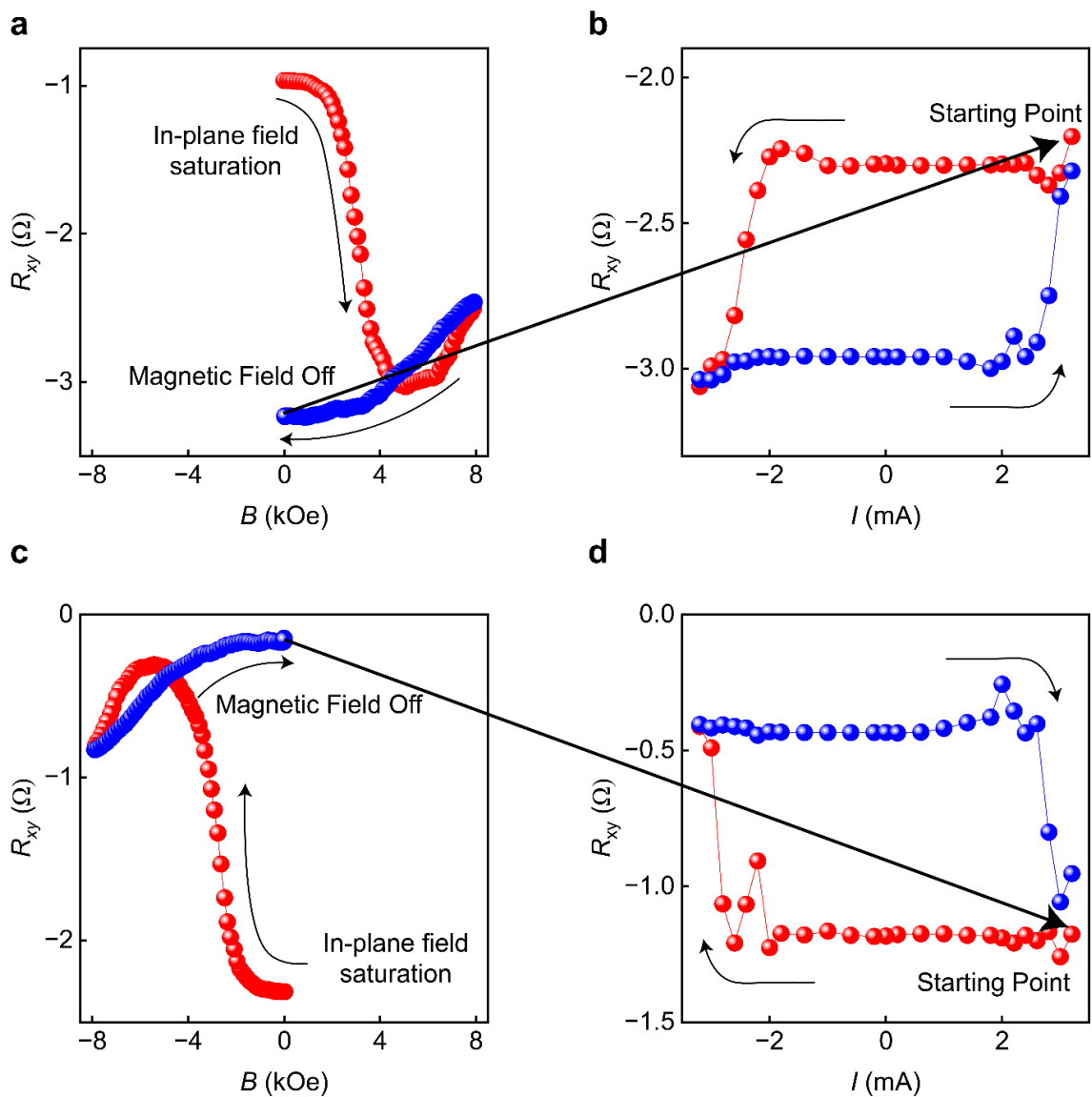

**Figure S8.** Polarity changes in Device 1, which were used in Figure 4a. a) 8 kOe magnetic field was applied and then turned off. b) Starting from 2.8 mA, the current decreased in steps of 0.2 mA to -2.8 mA and then increased back to 2.8 mA. c) Afterwards, the -8 kOe magnetic field was applied and turned off. d) Starting from 2.8 mA, a current sweep was performed.

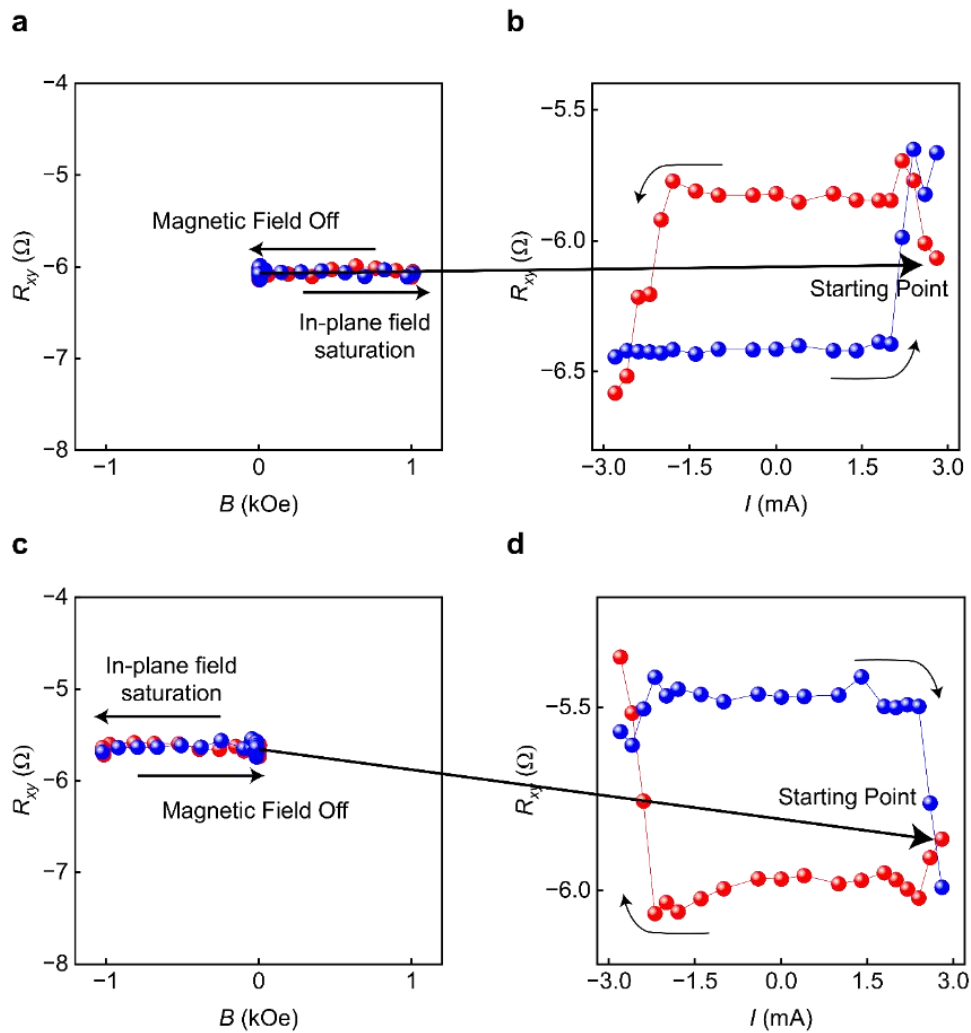

**Figure S9**. Polarity changes in Device 4 due to the application of an in-plane magnetic field. a) 1 kOe magnetic field was applied and then turned off. b) Starting from 2.8 mA, the current decreased in steps of 0.2 mA to -2.8 mA and then increased back to 2.8 mA. c) Afterwards, -1 kOe magnetic field was applied and then turned off. d) Starting from 2.8 mA, a current sweep was performed. The small in-plane magnetic field of 1 kOe does not change FGT's magnetisation in (a) and (c), so the in-plane magnetic field mainly affects the interface magnetisation rather than FGT's magnetisation, eventually leading to the switching polarity change in (b) and (d).

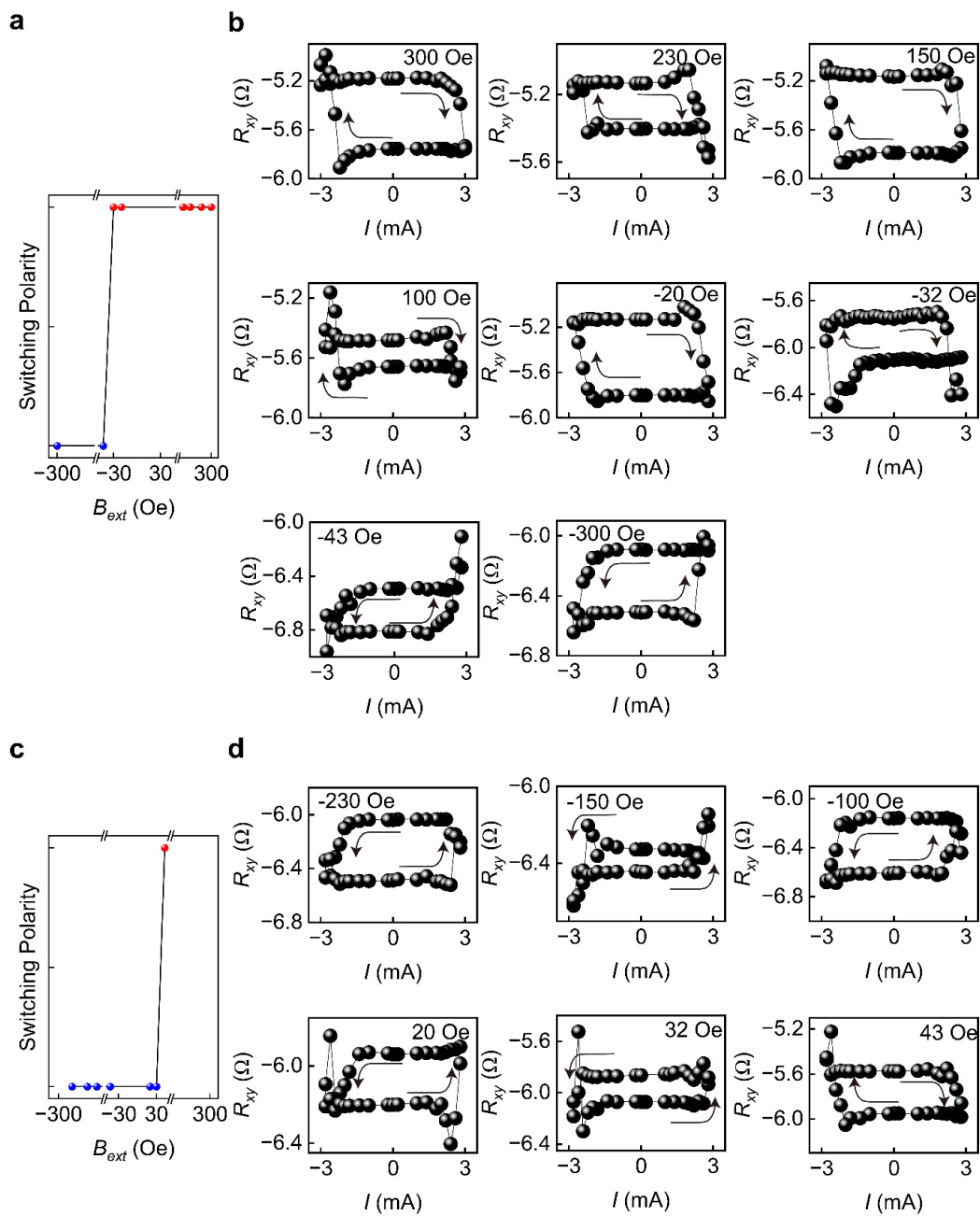

**Figure S10.** All $R(I)$ loops recorded for Figure 4b. a) An in-plane magnetic field was applied from 300 Oe to -300 Oe. b) Magnetic-field-free switching loops indicated in (a). c) An in-plane magnetic field was applied from -230 Oe to 43 Oe. d) Magnetic-field-free switching loops indicated in (c).

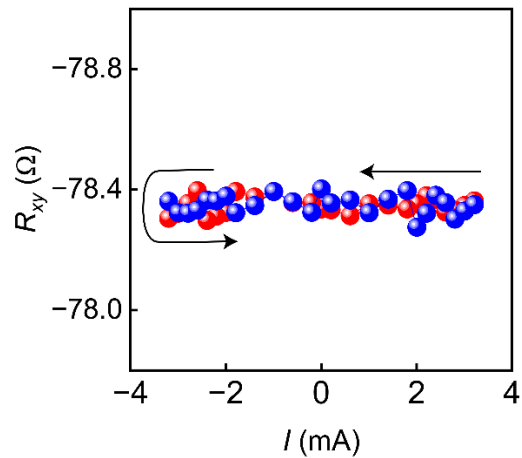

**Figure S11.** Hall resistance $R_{xy}$ of the FGT/STO device without an Ar-milled surface as a function of writing current $I$ without magnetic field at 120 K. No field-free switching is overserved in this comparison device. It supports our claim that the conducting (Ar-milled) STO surface interfacing with FGT is important for field-free switching. This observation also points to the possibility that Ar-etching and interfacing with FGT produces the in-plane interface magnetisation.

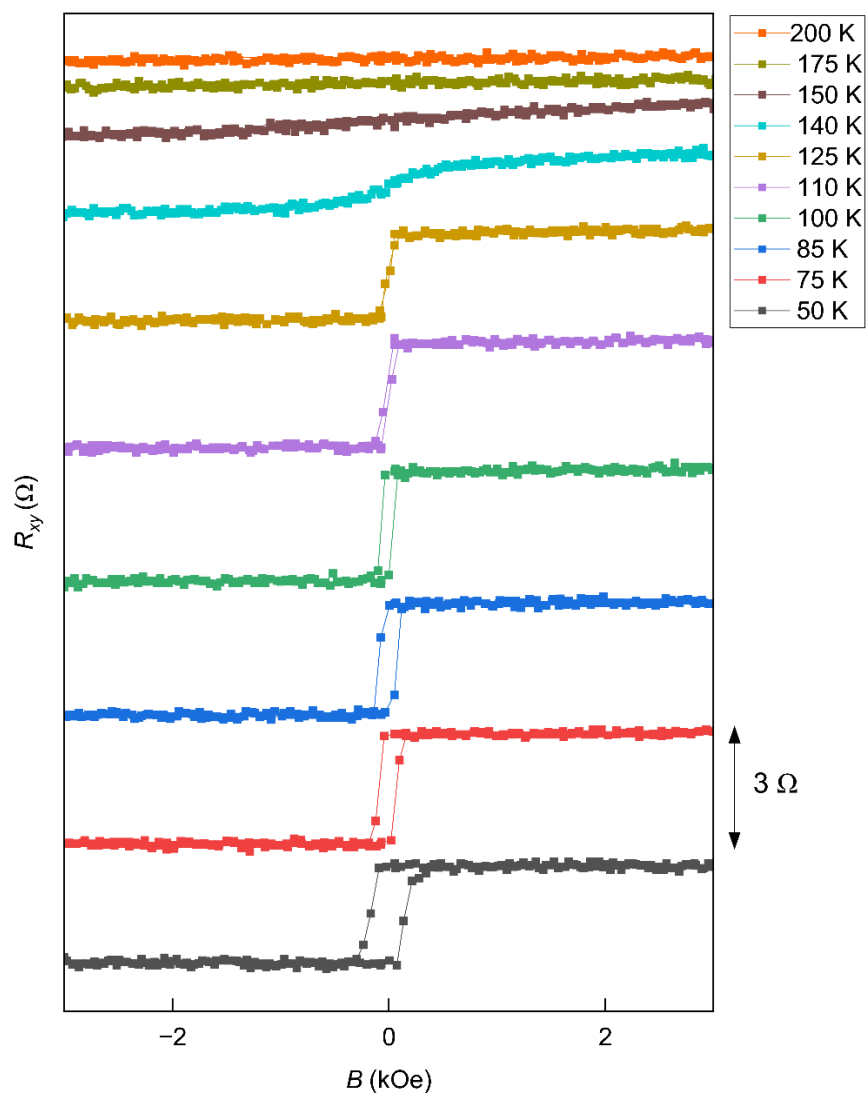

**Figure S12.** Anomalous Hall loops of the FGT/STO device without an Ar-milled surface at various temperatures.

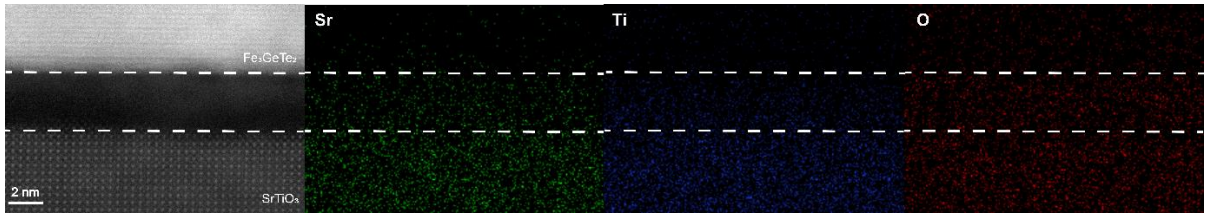

**Figure S13.** Cross-section STEM image of $Fe_3GeTe_2$/$SrTiO_3$ interface and EDS mapping image of Sr, Ti and O. The area enclosed by the dotted line is the plasma-treated $SrTiO_3$. We can see a significant contrast difference compared to the crystalline $SrTiO_3$. Additionally, contrast differences are observed in the EDS results for Sr, Ti, and O. The Sr and Ti vacancies may be able to create the magnetism[4], indicating that the origin of the in-plane magnetisation most likely comes from Ar-milled $SrTiO_3$ surface defects interfacing with FGT.

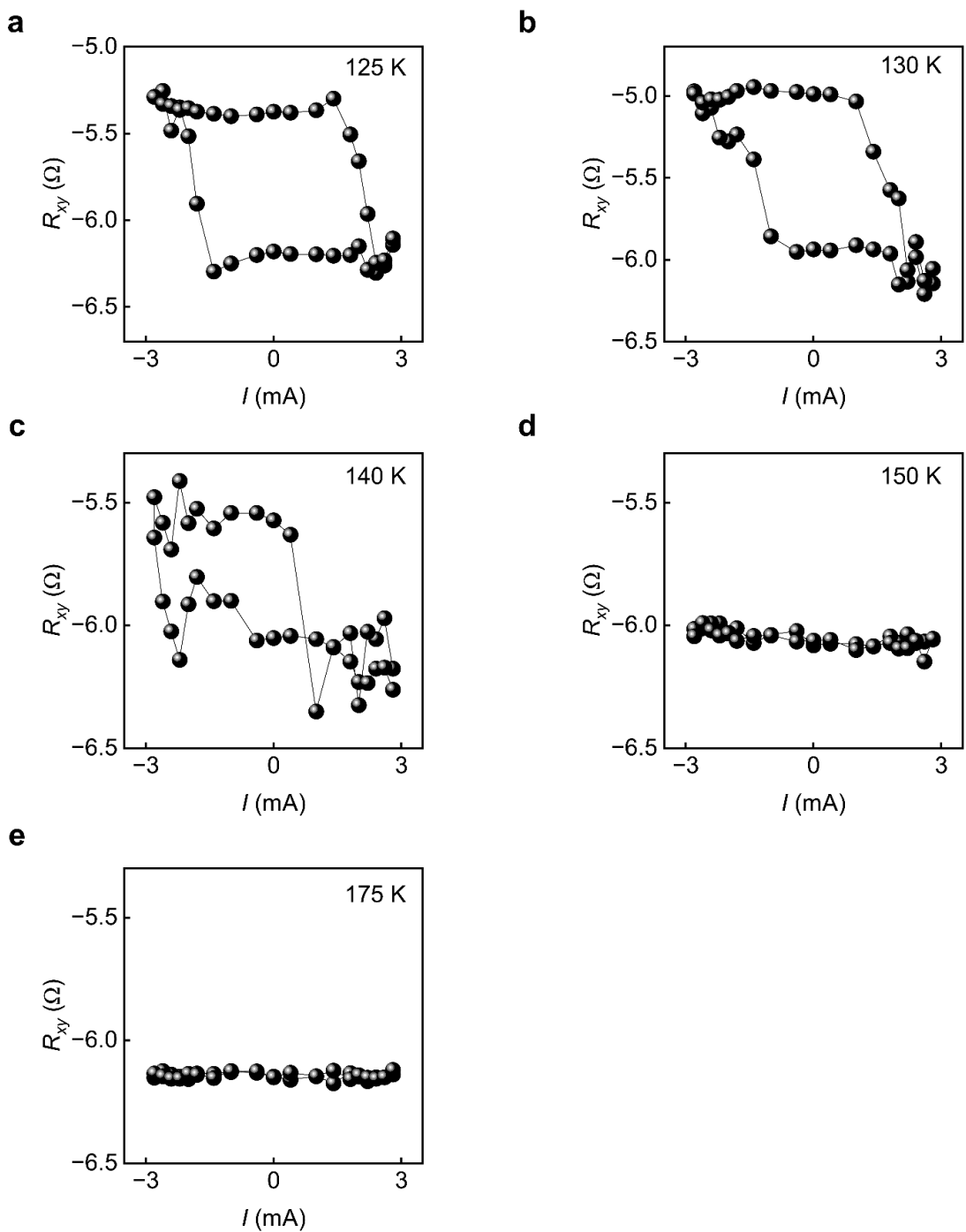

**Figure S14.** $R(I)$ loops of Device 4 at various temperatures. Raw data of $R_{xy}$ as a function of writing current $I$, without external magnetic field at 125 K (a), 130 K (b), 140 K (c), 150 K (d) and 175 K (e).

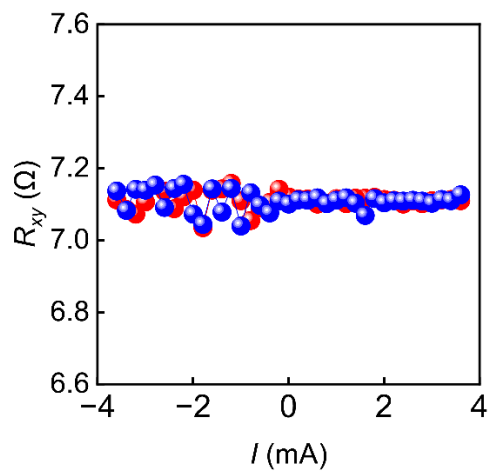

**Figure S15.** Hall resistance $R_{xy}$ of the FGT/STO device with a 22 nm-thick FGT, as a function of writing current $I$ without magnetic field at 120 K. No field-free switching is observed in this case, highlighting the interfacial effect in our field-free switching phenomena. Based on these field-free switching experiments, we think the suitable thickness range of FGT is around 9-15 nm for our FGT/STO field-free switching devices.

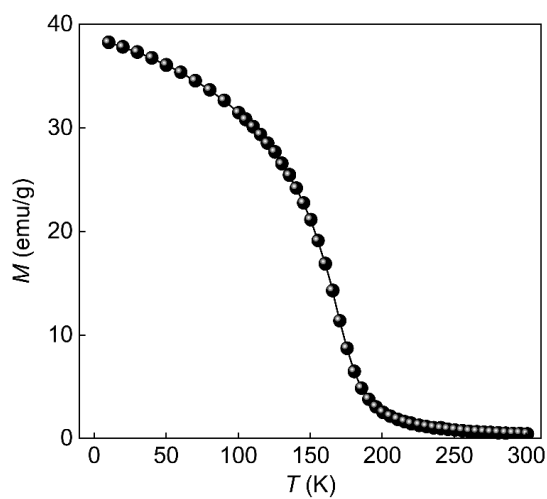

**Figure S16.** Temperature dependence of the magnetisation for bulk FGT under a 5 kOe magnetic field along the *c*-axis. It indicates a bulk $T_c$ of around 170 K for our FGT single crystals.

.